\documentclass{llncs}
\usepackage[english]{babel}
\usepackage[T1]{fontenc}
\usepackage{ae,aecompl}
\usepackage{marvosym}
\usepackage{amsmath}
\usepackage{graphicx}
\usepackage[frozencache]{minted}
\usepackage{tcolorbox}
\usepackage{todonotes}
\usepackage{fancyhdr}
\usepackage[bookmarks,bookmarksopen,bookmarksdepth=3]{hyperref}
\setminted[haskell]{escapeinside=@@}
\newcommand{\hs}{\mintinline{haskell}}
\newcommand{\subhs}{\mintinline[fontsize=\small]{haskell}}
\makeatletter
  \defineshorthand[english]{"/}{\babelhyphen{nobreak}}
  \addto\extrasenglish{
    \languageshorthands{english}
    \useshorthands{"}
  }
\makeatother

\begin{document}
%
%
\pagestyle{headings}

\mainmatter              
\title{$\!\!\!$Formal Verification of Spacecraft Control Programs$\!\!\!$ Using a Metalanguage for State Transformers}
\titlerunning{Formal verification of spacecraft control programs}

\author{Andrey Mokhov\textsuperscript{1(\Letter)},
Georgy Lukyanov\textsuperscript{1},
Jakob Lechner\textsuperscript{2}}
\authorrunning{Andrey Mokhov, Georgy Lukyanov, Jakob Lechner}
\institute{\vspace{-2.5mm}\textsuperscript{1}Newcastle University, UK~~~~~\textsuperscript{2}RUAG Space Austria GmbH\vspace{0.5mm}\\
\raisebox{-0.3mm}{\scalebox{1.3}{\Letter}}~\href{mailto:andrey.mokhov@ncl.ac.uk}{\textsf{andrey.mokhov@ncl.ac.uk}}\vspace{-2mm}}

\maketitle

\thispagestyle{fancy}
\renewcommand{\headrulewidth}{0.2pt}
\chead{Under review, feedback is sought}

\begin{abstract}
Verification of functional correctness of control programs is an essential task
for the development of space electronics; it is difficult and time-consuming and
typically outweighs design and programming tasks in terms of development hours.
We present a verification approach designed to help spacecraft engineers reduce
the effort required for formal verification of low-level control programs executed
on custom hardware. The approach uses a metalanguage to describe the semantics
of a program as a state transformer, which can be compiled to multiple targets
for testing, formal verification, and code generation. The metalanguage itself
is embedded in a strongly-typed host language (Haskell), providing a way to prove
program properties at the type level, which can shorten the feedback loop and
further increase the productivity of engineers.

\hspace{4mm}The verification approach is demonstrated on an industrial case study.
We present REDFIN, a processing core used in space missions, and its formal
semantics expressed using the proposed metalanguage, followed by a detailed
example of verification of a simple control program.
\vspace{-3mm}
\end{abstract}

\section{Introduction}
\vspace{-1mm}
Software bugs play a major role in the history of spacecraft
accidents~\cite{Leveson2004}. There are recorded cases of mission-ending bugs
that would have been difficult to prevent (e.g. caused by concurrency
or updates) but also plain integer overflows~\cite{bug-rocket} and incorrect unit
conversion~\cite{NASA:1999:Mars}, which should have been eradicated long ago.

Alas, there is no silver bullet. Testing is supported by mature methodologies
and frameworks, but does not provide the full correctness guarantee. General-purpose
strongly-typed languages can be used to eliminate important classes of bugs, but
are less familiar to software engineers and are often not suitable for highly
resource-constrained microarchitectures used in space electronics. Formal modelling
methods provide a systematic approach for developing complex systems in a
correct-by-construction manner, but they are still at the bleeding-edge of
computing science and can be difficult to apply to real-life systems. This paper
combines known formal verification and programming languages techniques and presents
a formal verification approach for simple control tasks, such as satellite power
management, which are executed on a real processing core used in space missions.
We believe the presented ideas are transferable to other domains with similar
safety and resource requirements, e.g. biomedical applications.
\begin{figure}
\centerline{\includegraphics[scale=0.45]{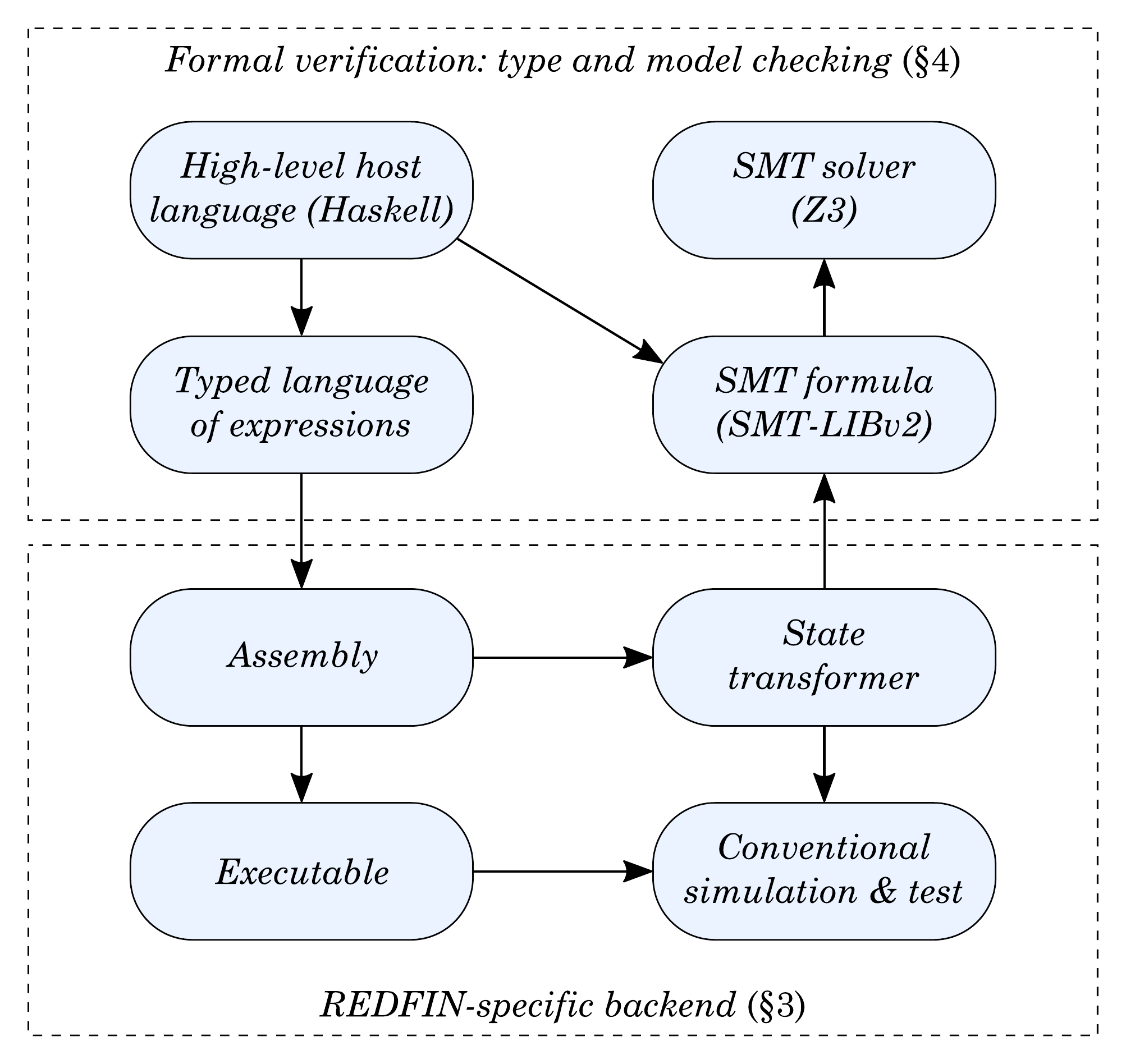}}
\vspace{-3mm}
\caption{Overview of the presented formal verification approach.\label{fig-overview}}
\vspace{-5mm}
\end{figure}

Fig.~\ref{fig-overview} shows an overview of the presented approach. The bottom
part corresponds to conventional code generation and simulation, where
REDFIN\footnote{REDFIN stands for `REDuced instruction set for Fixed-point \&
INteger arithmetic'. This instruction set and the corresponding processing core
were developed by RUAG Space Austria GmbH for space missions.
See~\S\ref{sec-redfin} for more details.} assembly language is executed by
simulating the effect of each instruction on the state of the processor and memory.
The corresponding \emph{state transformer} is typically implicit and intertwined
with the rest of the simulation infrastructure. The main idea of our approach is
to represent the state transformer explicitly so that it can be symbolically
manipulated and used not only for simulation but also for formal verification by
compiling it into an SMT formula. We can then~use~an~SMT solver, e.g.
Z3~\cite{de2008z3}, to verify that the state transformer of a given program
satisfies certain properties, for example, that integer overflow cannot occur
regardless of input parameters and that the program always terminates within
given time.

By embedding the state transformer metalanguage in Haskell we can readily
implement compilers from higher-level \emph{typed} languages to untyped assembly,
eradicating incorrect number and unit conversion bugs. As shown at the top of
Fig.~\ref{fig-overview}, engineers can write high-level control
programs for the REDFIN architecture directly in a small subset of Haskell. These
high-level programs can be used for type-safe code generation and as executable
specifications of intended functionality for the purposes of program synthesis and
equivalence checking. 

We first introduce the REDFIN processing core (\S\ref{sec-redfin}), and then
describe and discuss the presented approach
(\S\ref{sec-transformer}-\S\ref{sec-discussion}). Related work is reviewed
in~\S\ref{sec-related}.

\clearpage

\vspace{-3mm}
\section{The REDFIN overview\label{sec-redfin}}
\vspace{-3mm}
For many spacecraft subsystems integrated circuits are required to perform control
tasks or simple data processing. Typically, these integrated circuits are realised
with FPGAs (Field Programmable Gate Arrays) due to their flexibility and lower
costs compared to ASIC (Application-Specific Integrated Circuit) development \&
fabrication. Since FPGAs can be used to implement arbitrary circuit functions
including processor cores, it is possible to perform tasks both in hardware and
in software. However, modern space-qualified FPGAs, which can withstand radiation
in Earth orbit or deep space, have a limited amount of programmable resources.
Therefore, it is often not feasible to implement a fully-fledged processor system
in such an FPGA next to the mission-specific circuitry.

The REDFIN instruction set was developed to address this issue and, more
specifically, to meet the following goals: (i) simple instruction set to
achieve a small hardware footprint, (ii) reduced complexity to support formal
verification of programs, and (iii) deterministic behaviour for real-time
applications.

\vspace{-3mm}
\subsection{REDFIN instruction set and microarchitecture}
\vspace{-1mm}

The instruction set architecture offers a configurable bit width for the
data path, ranging from 8 to 64 bits. Instruction words have a fixed width of 16
bits. The instruction set is based on a register-memory architecture, i.e.
instructions can fetch their operands from registers as well as directly from the
memory. This architecture favours a small register set, which minimises the hardware
footprint of the processing core. Furthermore, the number of instructions in a
program is typically smaller in comparison to traditional load/store architectures
where all operands have to be transferred to registers before any operations can
be performed. There are 47 instructions of the following types:

\vspace{-2.5mm}
\begin{itemize}
\item{Load/store instructions for moving data words between general- and
special-purpose registers and the memory, as well as load of immediate values.}
\item{Arithmetic operations for integer and fixed-point numbers. In the latter
case the number of fractional bits can be adjusted by a processor register.}
\item{Bitwise logical and shift operations.}
\item{Control flow instructions and associated comparison operations.}
\item{Bus access instructions for read \& write operations on an AMBA AHB bus.}
\end{itemize}
\vspace{-2mm}

The REDFIN processing core fetches instruction and data words from a small and fast
on-chip SRAM. This only allows for execution of simple programs, however, it also
eliminates the need to implement caches and thus removes a source of non-determinism of
conventional processors. Since computing performance is not one of the main goals, the
processor core is non-pipelined and therefore does not need to resolve data or control
hazards or perform any form of speculative execution. These properties greatly simplify
worst case execution time analysis.

\vspace{-3mm}
\subsection{Requirements for formal verification}
\vspace{-1mm}

Verification of \emph{functional correctness} of REDFIN programs, as defined by a
requirement specification, clearly is an essential task for the development of space
electronics. There are also important \emph{non-functional requirements}, such as
worst case execution time and energy consumption, which rely on the implementation
guarantees provided by the processing core. In order to reduce verification complexity,
the REDFIN core only allows to execute a single subroutine whose execution is triggered
by a higher-level controller in the system. The implementation also guarantees
that concurrent bus accesses to the processor registers or memory do not affect
the subroutine execution time. Furthermore, the processor does not implement
interrupt handling. All these measures are taken in order to provide real-time
subroutine execution guarantees and make the verification of non-functional
properties feasible within the presented verification framework.

Despite these restrictions the REDFIN core has already proven its effectiveness for
simple control tasks and arithmetic computations as part of an antenna pointing unit
for satellites. Nevertheless, verification can be difficult and time-consuming,
even for small and simple programs. Verification activities, following engineering
standards for space electronics, typically outweigh programming and design tasks by a
factor of two in terms of development hours. Usually verification is performed via
program execution on an instruction set simulator or a hardware model of the processor.
Manually deriving test cases from the specification is cumbersome and error-prone
and simulation times can become prohibitively long with a large number of tests that
are often needed to reach the desired functional and code coverage. Formal verification
methods can prove that a program satisfies certain properties for all possible
test cases and are therefore immensely valuable for completing the verification
with superior efficiency and quality.

\vspace{-3mm}
\section{State transformer\label{sec-transformer}}
\vspace{-2mm}
In this section we formally define the REDFIN microarchitecture and express the
semantics of the instruction set as an explicit and symbolic state transformer.

\vspace{-3mm}
\subsection{The REDFIN microarchitecture state}
\vspace{-1mm}

The main idea of the presented approach is to use an explicit state"/transformer
semantics of the REDFIN microarchitecture. The state space of the entire
processing core is a Cartesian product of state spaces of every component:

\vspace{-5mm}
\[
S=\{(r, m, ic, ir, p, f, c) : r \in R, m \in M, ic \in A, ir \in I, p \in P, f \in F, c \in C\},
\]

\noindent
where $R$ is the set of register bank configurations;
$M$ is the memory state space;
$A$~is the set of instruction addresses (the instruction counter $ic$ stores the
address of the current instruction);
$I$ is the set of instruction codes (the instruction register $ir$ stores the
code of the current instruction);
$P$ is the set of programs;
$F$ is the set of the flag register configurations; and
$C$ is the set of clock values.

Fig.~\ref{fig-types} shows the translation of the above into Haskell types. Note
that the types are not parameterised: recall that REDFIN is parameterised, e.g.
the data width can be chosen depending on mission requirements, whereas we use fixed
64-bit data path for the sake of simplicity. The chosen names are self-explanatory,
for example, the data type \hs{State} directly corresponds to the set of states~$S$.
We defined \hs{SymbolicValue} and \hs{SymbolicArray} type constructors on top of
Levent Erkok's symbolic verification library SBV~\cite{SBV}, which we use as the
SMT translation and verification frontend. In principle, any other SMT frontend
can be used, but to the best of our knowledge, SBV is the most mature SMT library
available for Haskell. We briefly overview all \hs{State} components below.

\begin{figure}[t]
\vspace{-2mm}
\begin{minted}{haskell}
      data State = State { registers           :: RegisterBank
                         , memory              :: Memory
                         , instructionCounter  :: InstructionAddress
                         , instructionRegister :: InstructionCode
                         , program             :: Program
                         , flags               :: Flags
                         , clock               :: Clock }
\end{minted}
\vspace{-4mm}
\begin{minted}{haskell}
      type Register           = SymbolicValue Word2
      type Value              = SymbolicValue Int64
      type RegisterBank       = SymbolicArray Word2 Int64
\end{minted}
\vspace{-4.5mm}
\begin{minted}{haskell}
      type MemoryAddress      = SymbolicValue Word8
      type Memory             = SymbolicArray Word8 Int64
\end{minted}
\vspace{-4.5mm}
\begin{minted}{haskell}
      type InstructionAddress = SymbolicValue Word8
      type InstructionCode    = SymbolicValue Word16
      type Program            = SymbolicArray Word8 Word16
\end{minted}
\vspace{-4.5mm}
\begin{minted}{haskell}
      data Flag               = Condition | Overflow | Halt | ...
      type Flags              = SymbolicArray Flag Bool
\end{minted}
\vspace{-4.5mm}
\begin{minted}{haskell}
      type Clock              = SymbolicValue Word64
\end{minted}
\vspace{-5mm}
\caption{Basic types for encoding REDFIN microarchitecture in Haskell.\label{fig-types}}
\vspace{-4mm}
\end{figure}

\vspace{-4.5mm}
\subsubsection{Data values, registers and memory}
64-bit data values (\hs{Int64}) are stored in registers and memory. There are~4
registers (addressed by \hs{Word2}) and 256 memory locations (addressed by
\hs{Word8}). Their content is represented by symbolic arrays that can be
accessed via SBV's functions \hs{readArray} and \hs{writeArray}.

\vspace{-4.5mm}
\subsubsection{Instructions and programs}
REDFIN instructions are represented by 16-bit \hs{InstructionCode}, whose 6
leading bits contain the instruction opcode, and the remaining 10 bits are
allocated for various instruction arguments. The \hs{Program} is a symbolic
array mapping 8-bit instruction addresses to instruction codes.

\vspace{-4.5mm}
\subsubsection{Status flags and clock}
The microarchitecture stores execution status flags to support conditional
branching, track integer overflow, and terminate the program, as captured by the
data type \hs{Flag} (we omit a few other flags for brevity). The flag register is a
symbolic map from flags to Boolean values. The \hs{Clock} is a 64-bit counter
incremented on each clock cycle. Status flags and the clock are used for
diagnostic, formal verification and worst-case execution time analysis.

\vspace{-3mm}
\subsection{Instruction and program semantics}
\vspace{-1mm}
We can now define the formal semantics of REDFIN instructions and programs in terms
of a \emph{state transformer} $T : S \rightarrow S$, i.e. a function that maps
states to states. We distinguish between instructions and programs by using
Haskell's list notation, for example, $T_{\subhs{nop}}$ is the semantics of the
instruction $\hs{nop} \in I$, whereas $T_{\subhs{[}\subhs{nop}\subhs{]}}$ is the
semantics of the single-instruction program $\hs{[}\hs{nop}\hs{]} \in P$.


\textbf{Definition (program semantics):} The semantics of a program $p \in P$
is inductively defined as follows:
\vspace{-2mm}

\begin{itemize}
    \item The semantics of the \emph{empty program} $\hs{[}\hs{]} \in P$ coincides with
    the semantics of the instruction \hs{nop} and is the identity state transformer:
    $T_{\subhs{[}\subhs{]}} = T_{\subhs{nop}} = \hs{id}$.
    \vspace{0.5mm}
    \item The semantics of the \emph{single-instruction program} $\hs{[}\hs{i}\hs{]} \in P$
    is a composition of three state transformers: (i) fetching the instruction from
    the program memory (denoted by $T_\textit{fetch}$), (ii) incrementing the
    instruction counter ($T_\textit{inc}$), and (iii) the state transformer
    of the instruction itself ($T_{\subhs{i}}$), that is:
    \vspace{-1.5mm}
    \[
    \begin{array}{lcl}
    T_\textit{fetch} & = & (r, m, ic, ir, p, f, c) \mapsto (r, m, ic, p[ic], p, f, c + 1)\vspace{0.7mm}\\
    T_\textit{inc} & = & (r, m, ic, ir, p, f, c) \mapsto (r, m, ic + 1, ir, p, f, c)\vspace{0.7mm}\\
    T_{\subhs{[}\subhs{i}\subhs{]}} & = & T_{\subhs{i}} \circ T_\textit{inc} \circ T_\textit{fetch}.\\
    \end{array}
    \]
    \vspace{-1.5mm}
    \item The semantics of the \emph{composite program} $\hs{i}\hs{:}\hs{p} \in P$,
    where the operator~\hs{:}~prepends an instruction $\hs{i} \in I$ to a program
    $\hs{p} \in P$, is defined as $T_{\subhs{i}\subhs{:}\subhs{p}} = T_{\hs{p}} \circ T_{\subhs{[}\subhs{i}\subhs{]}}$.
    \vspace{-1mm}
\end{itemize}

\noindent
We represent state transformers in Haskell using the \emph{state monad}, a
classic approach to emulating mutable state in a purely functional programming
language~\cite{wadler1990comprehending}. We call our state monad~\hs{Redfin} and
define it\footnote{\mbox{A generic version of this monad is available in standard module
\hs{Control.Monad.State}.}} as follows:

\vspace{-0.5mm}
\begin{minted}{haskell}
data Redfin a = Redfin { transform :: State -> (a, State) }
\end{minted}
\vspace{-0.5mm}

\noindent
Every computation with the return type~\hs{Redfin} \hs{a} yields a value of type~\hs{a}
and possibly alters the \hs{State} of the REDFIN microarchitecture. As an example,
below we express the state transformer $T_\textit{inc}$ using the \hs{Redfin} monad.

\vspace{-0.5mm}
\begin{minted}{haskell}
incrementInstructionCounter :: Redfin ()
incrementInstructionCounter = Redfin $ \current -> ((), next)
  where
    next = current { instructionCounter = instructionCounter current + 1 }
\end{minted}
\vspace{-0.5mm}

\noindent
In words, the state transformer looks up the value of the \hs{instructionCounter}
in the \hs{current} state and replaces it in the \hs{next} state with the
incremented~value. The type \hs{Redfin ()} indicates that the computation does not
produce any value as part of the state transformation. Such computations directly
correspond to REDFIN programs and can be composed using the operator \hs{>>}.
For example, \hs{fetchInstruction}~\hs{>>}~\hs{incrementInstructionCounter} is
the state transformer $T_\textit{inc} \circ T_\textit{fetch}$ assuming that
\hs{fetchInstruction} corresponds to $T_\textit{fetch}$. We can also use
Haskell's powerful \hs{do}-notation to compose computations:

\vspace{-1mm}
\begin{minted}{haskell}
readInstructionRegister :: Redfin InstructionCode
readInstructionRegister = Redfin $ \s -> (instructionRegister s, s)
\end{minted}
\vspace{-3mm}
\begin{minted}{haskell}
executeInstruction :: Redfin ()
executeInstruction = do fetchInstruction
                        incrementInstructionCounter
                        instructionCode <- readInstructionRegister
                        decodeAndExecute instructionCode
\end{minted}

\noindent
Here \hs{readInstructionRegister} extracts the instruction code from the current
state \emph{without modifying it}. This function is used in \hs{executeInstruction},
which defines the semantics of the REDFIN execution cycle. We omit the definition of
\hs{decodeAndExecute} for brevity: it is a case analysis of 47 opcodes that returns
the matching instruction. We introduce several interesting instructions below.

\vspace{-4.5mm}
\subsubsection{Halting the processor}
If the~\hs{halt} instruction is encountered, the processor sets the flag~\hs{Halt},
thereby stopping the execution of the current subroutine until a new one is
started by a higher-level system controller that resets \hs{Halt}. The auxiliary
function \hs{writeFlag} is used to do the actual flag modification.

\vspace{-1mm}
\begin{minted}{haskell}
halt :: Redfin ()
halt = writeFlag Halt true
\end{minted}
\vspace{-3mm}
\begin{minted}{haskell}
writeFlag :: Flag -> SymbolicValue Bool -> Redfin ()
writeFlag flag value = Redfin $ \s -> ((), s')
  where s' = s { flags = writeArray (flags s) (flagId flag) value }
\end{minted}
\vspace{-1mm}

\vspace{-5mm}
\subsubsection{Arithmetics}
As a more involved example, consider the semantics of the instruction \hs{abs}.
It reads a register and writes back the absolute value of its
contents\footnote{We use \hs{Prelude.abs} to distinguish between the instruction
and the function from the standard library \hs{Prelude}; \hs{fmap} applies
\hs{Prelude.abs} to the result of \hs{readRegister}.}.
The semantics accounts for the potential integer overflow that leads to the
\emph{negative resulting value} when the input is $-2^{63}$ (REDFIN
uses the two's complement signed number representation). The overflow is flagged
by setting~\hs{Overflow}.

\vspace{-1.5mm}
\begin{minted}{haskell}
abs :: Register -> Redfin ()
abs rX = do
    state  <- readState
    result <- fmap Prelude.abs (readRegister rX)
    let (_, overflowState) = transform (writeFlag Overflow true) state
    writeState $ ite (result .< 0) overflowState state
    writeRegister rX result
\end{minted}
\vspace{-1mm}

\noindent
Here, SBV's symbolic \emph{if-then-else} operation~\hs{ite} is used to \emph{merge}
two possible next states, one of which has the \hs{Overflow} flag set. We use
auxiliary functions \hs{readRegister}, \hs{writeRegister}, \hs{readState} and
\hs{writeState} --- simple state transformers defined similarly to
\hs{readInstructionRegister} and \hs{writeFlag}.

\vspace{-4.5mm}
\subsubsection{Conditional branching}
As an example of a control flow instruction, consider the conditional branching
instruction \hs{jmpi_ct}, which tests the~\hs{Condition} flag, and adds the
provided offset to the instruction counter if the flag is set.

\vspace{-1mm}
\begin{minted}{haskell}
jmpi_ct :: SymbolicValue Int8 -> Redfin ()
jmpi_ct offset = do ic <- readInstructionCounter
                    condition <- readFlag Condition
                    let ic' = ite condition (ic + offset) ic
                    writeInstructionCounter ic'
\end{minted}
\vspace{-1mm}

\noindent
After working through the above examples, it is worth noting that we use our
Haskell encoding of the state transformer as a~\emph{metalanguage}. We are
operating the REDFIN core as a puppet master, using external
\pagebreak
meta-notions of addition,
comparison and let"/binding. From the processor's
point of view, we have infinite memory and act instantly, which gives us unlimited
modelling power. For example, we can run a simulation of the processor environment
in an external tool and feed its result to \hs{writeRegister} as if it was
obtained in a single clock cycle.

\vspace{-1.5mm}
\subsection{Symbolic simulation}
\vspace{-0.5mm}
Having defined the semantics of REDFIN instructions and programs, we can
implement symbolic simulation of the processor:

\vspace{-1mm}
\begin{minted}{haskell}
simulate :: Int -> State -> State
simulate steps state
    | steps <= 0 = state
    | otherwise  = ite halted state (simulate (steps - 1) nextState)
  where
    halted    = readArray (flags state) (flagId Halt)
    nextState = snd $ transform executeInstruction state
\end{minted}
\vspace{-1mm}

\noindent
The function takes a number of simulation steps $N$ and an initial symbolic
state of the processor as input, and executes $N$ instructions using the
previously defined \hs{executeInstruction} function. In each \hs{state} we need
to merge two possible futures depending on the value of the \hs{Halt} flag:
(i) continue the simulation starting from the \hs{nextState} if the flag is not
set, and (ii) remain in the current \hs{state} if the flag is set, since in
this case the processor must remain idle.

Symbolic simulation is very powerful. It allows us to formally verify properties
of REDFIN programs by fixing some parts of the state to constant values (for
example, the program code), and then checking assertions on the resulting values of
the symbolic part of the state. This will be discussed in the next
section~\S\ref{sec-verification}.

\vspace{-1.5mm}
\section{Formal verification\label{sec-verification}}
\vspace{-1.5mm}
This section presents the formal verification framework developed on top of
the REDFIN semantic core (\S\ref{sec-transformer}) demonstrating the following
steps of the workflow:

\vspace{-1mm}
\begin{itemize}
    \item Develop programs either in low-level REDFIN assembly, or in a high-level
    and statically type-checked expression language embedded in Haskell.
    \item Initialise, execute and test REDFIN programs on concrete input values.
    \item Formulate and refine functional correctness and worst case execution time
    properties in the SBV property specification language.
    \item Verify the properties with an SMT solver.
    \item Receive the verification results (e.g. counterexamples) for analysis.
\end{itemize}

\noindent
As our running example, consider the following simple spacecraft control task.

\vspace{1mm}
\begin{tcolorbox}
\vspace{-1mm}
Let $t_1$ and $t_2$ be two different time points (measured in ms),
and $p_1$ and $p_2$ be two power values (measured in mW).
Calculate the estimate of the total energy consumption during this period
using linear approximation, rounding down to the nearest integer:
\vspace{-3mm}
\[
\textit{energyEstimate}(t_1, t_2, p_1, p_2) = \left\lfloor \frac{|t_1 - t_2| * (p_1 + p_2)}{2} \right\rfloor.
\]
\vspace{-5.5mm}
\end{tcolorbox}

\noindent
This task looks too simple, but in fact it has a few critical pitfalls that,
if left unattended, may lead to the failure of the whole space mission. Examples
of subtle bugs in seemingly simple programs leading to a catastrophe include 64"/bit
to 16"/bit number conversion overflow causing the destruction of Ariane~5
rocket~\cite{bug-rocket} and the loss of NASA's Mars climate orbiter due to incorrect
units of measurement conversion~\cite{NASA:1999:Mars}. Let us develop and verify
a REDFIN program for this task.

\vspace{-5mm}
\subsubsection{Writing the program}
We can write programs either directly in the untyped REDFIN assembly, or in a
typed higher-level expression language. Both have their advantages: the former
allows engineers to hand-craft highly optimised programs under tight resource
constraints, and the latter brings type-safety and faster prototyping. Our first
prototype therefore uses the high-level approach.

Using Haskell's polymorphism, we can define an expression that can
be used both as a Haskell function and a high-level REDFIN expression:

\vspace{-1mm}
\begin{minted}{haskell}
energyEstimate :: Integral a => a -> a -> a -> a -> a
energyEstimate t1 t2 p1 p2 = abs (t1 - t2) * (p1 + p2) `div` 2
\end{minted}
\vspace{-1mm}

\noindent
We implement the energy estimation program by embedding the~\hs{energyEstimate}
expression into a REDFIN~\hs{Script}. Due to the lack of space we omit the
implementation of \hs{Script}, but one can think of it as a restricted version
of the \hs{Redfin} state transformer, which we use to write \emph{programs that
can manipulate the processor state only by executing instructions}, e.g. the
only way to set the \hs{Overflow} flag is to execute an arithmetic instruction
that actually has an overflow.

\vspace{-1mm}
\begin{minted}[linenos]{haskell}
energyEstimateHighLevel :: Script
energyEstimateHighLevel = do
    let t1    = read (IntegerVariable 0)
        t2    = read (IntegerVariable 1)
        p1    = read (IntegerVariable 2)
        p2    = read (IntegerVariable 3)
        temp  = Temporary 4
        stack = Stack 5
    compile r0 stack temp (energyEstimate t1 t2 p1 p2)
    halt
\end{minted}
\label{energyEstimateHighLevel}
\vspace{-1mm}

\noindent
Here the type \hs{IntegerVariable} is used to statically distinguish between integer
and fixed-point numbers, \hs{Temporary} to mark temporary words, so they cannot
be mixed with inputs and outputs, and \hs{Stack} to denote the location of the
stack pointer. The~\hs{let} block declares six adjacent memory addresses: four
input values $\{t_1, t_2, p_1, p_2\}$, a temporary word and a stack pointer.
The~\hs{compile} function call at line~9 performs the embedding of the high-level
\hs{energyEstimate} expression into the assembly language by translating it to a
sequence of REDFIN instructions. The first argument of the~\hs{compile} function
holds the register~\hs{r0} which contains the estimated energy value after the
program execution.

\vspace{-5mm}
\subsubsection{Simulating the program}
We run symbolic simulation for 100 steps, initialising the program and data
memory of the processor using the helper function \hs{boot}.

\begin{minted}{haskell}
main :: IO ()
main = do
    let dataMemory = [10, 5, 3, 5, 0, 100]
        finalState = simulate 100 (boot energyEstimateHighLevel dataMemory)
    printMemoryDump 0 5 (memory finalState)
    putStrLn $ "R0: " ++ show (readArray (registers finalState) r0)
\end{minted}
\vspace{-1mm}

\noindent
As the simulation result we get a \hs{finalState} value. We can inspect it by
printing relevant components: the values of the first six memory cells, and the
result of the computation located in the register~\hs{r0}. Note that the stack
pointer (cell 5) holds 100 as in the initial state, which means the stack is empty.

\vspace{-1mm}
\begin{minted}[frame=single, fontsize=\small]{text}
Memory dump: [10, 5, 3, 5, 5, 100]
R0: 20
\end{minted}
\vspace{-2mm}

Simulating programs with concrete input values is useful for diagnostic and test.
However, to formally verify their functional correctness, we need to inspect
every valid combination of input values, which can be done by an SMT solver.
This allows us to discover a trap hidden in our energy estimation program.

\vspace{-5mm}
\subsubsection{Verifying the program}
The project lead engineer defined a set of functional requirements for the
energy monitoring subsystem. The software engineering team received the
specification, implemented the energy monitoring subroutines, and started the
verification. One of the requirements is as follows.

\begin{tcolorbox}
\vspace{-1mm}
Assuming that values $p_1$ and $p_2$ are non"/negative integers, the energy
estimation subroutine must always return a non-negative integer value.
\vspace{-1mm}
\end{tcolorbox}

To check that the program complies with the requirement, we translate the
symbolic state transformer \hs{energyEstimateHighLevel} into an SMT formula,
and formulate the following theorem. Lines 13-14 extract the computed result and
the value of the flag \hs{Halt} from the \hs{finalState}. Lines 15-16 require that
the processor has halted, the result is equal to that computed by the high-level
Haskell expression \hs{energyEstimate} and is non"/negative.

\vspace{-1mm}
\begin{minted}[linenos, fontsize=\small]{haskell}
theorem = do
    -- Initialise symbolic variables:
    t1 <- forall "t1"
    t2 <- forall "t2"
    p1 <- forall "p1"
    p2 <- forall "p2"
    -- Constrain p1 and p2 to be non-negative:
    constrain $ p1 .>= 0 &&& p2 .>= 0
    -- Initialise the data memory with symbolic variables:
    let dataMemory = [t1, t2, p1, p2, 0, 100]
        finalState = simulate 100 (boot energyEstimateHighLevel dataMemory)
        result     = readArray (registers finalState) r0
        halted     = readArray (flags finalState) (flagId Halt)
    -- Specify the requirements as a property for verification:
    return $ halted &&& result .== energyEstimate t1 t2 p1 p2
                    &&& result .>= 0
\end{minted}

\noindent
The resulting SMT formula has 122 clauses and can be checked by Z3 in
3.0s\footnote{All reported results are obtained on a laptop with 2.90GHz Intel
Core i5-4300U processor, 8GB RAM (3MB cache), and the SMT solver Z3 version 4.5.1 (64-bit).}:

\vspace{-1.5mm}
\begin{minted}[frame=single]{text}
> proveWith z3 theorem
Falsifiable. Counter-example:
  t1 = 5190405167614263295 :: Int64
  t2 =                   0 :: Int64
  p1 =  149927859193384455 :: Int64
  p2 =  157447350457463356 :: Int64
\end{minted}
\vspace{-3mm}

The solver has found a counterexample demonstrating that the program does not
satisfy the above requirement. Indeed, the expression evaluates to a negative
value on the provided inputs due to an \emph{integer overflow}. The following
refined requirement is then provided by the project lead engineer:

\vspace{1mm}
\begin{tcolorbox}
\vspace{-1mm}
According to the spacecraft power system specification, $p_1$ and $p_2$ are
non"/negative integers that do not exceed 1W. The time is measured
from the mission start, hence $t_1$ and $t_2$ are non-negative and do not exceed
the time span of the mission, which is 30 years. Under these assumptions,
the energy estimation subroutine must return a non-negative integer value.
\vspace{-1mm}
\end{tcolorbox}

\noindent
The software engineering team modifies the time\footnote{We are not absolutely
precise here. We do not distinguish between regular and leap years, and use
a conservative upper bound of 366 days per year.} and power constraints:

\vspace{-1mm}
\begin{minted}{haskell}
constrain $ t1 .>= 0 &&& t1 .<= toMilliSeconds (30 :: Year)
constrain $ t2 .>= 0 &&& t2 .<= toMilliSeconds (30 :: Year)
constrain $ p1 .>= 0 &&& p1 .<= toMilliWatts   ( 1 :: Watt)
constrain $ p2 .>= 0 &&& p2 .<= toMilliWatts   ( 1 :: Watt)
\end{minted}
\vspace{-1mm}

\noindent
We rerun Z3 (now on 134 SMT clauses) and get the desired outcome in 4.8s:

\vspace{-1mm}
\begin{minted}[frame=single]{text}
> proveWith z3 theorem
Q.E.D.
\end{minted}
\vspace{-1mm}

\noindent
The refinement of the requirement has rendered the integer overflow impossible,
in particular, we can now be sure that \hs{abs} cannot be called with~$-2^{63}$
within the mission parameters. This kind of guarantee fundamentally requires
solving an SMT problem, even if it is done at the type level, e.g. using
\emph{refinement types}~\cite{vazou2014refinement}.

\vspace{-4mm}
\subsubsection{Checking program equivalence}
The statically typed high-level expression language is very convenient for
writing REDFIN programs, however, an experienced engineer can often find a way
to improve the resulting code. In some particularly resource-constrained situations,
a fully hand-crafted assembly code may be required. As an example, a direct
unoptimised translation of the \hs{energyEstimate} expression into assembly uses
79 instructions, most of them for \hs{Stack} manipulation.
On the other hand, it is not difficult to write a low-level assembly program that
computes the result using only 9 instructions as we demonstrate below.

To facilitate the development of verified hand"/crafted code, we use an SMT
solver to check the equivalence of REDFIN programs by verifying that they produce
the same output on all valid inputs. This allows an engineer to manually
optimise a given high-level prototype and have a guarantee that no bugs were
introduced in the process.

An optimised low-level energy estimation program has only 9 instructions:

\vspace{-1.5mm}
\begin{minted}{haskell}
energyEstimateLowLevel :: Script
energyEstimateLowLevel = do
    let { t1 = 0; t2 = 1; p1 = 2; p2 = 3 }
    ld r0 t1
    sub r0 t2
    abs r0
    ld r1 p1
    add r1 p2
    st r1 p2
    mul r0 p2
    sra_i r0 1
    halt
\end{minted}
\label{energyEstimateLowLevel}
\vspace{-1.5mm}

\noindent
Below we define the \hs{equivalence} check of this low"/level program with the
high-level program \hs{energyEstimateHighLevel} introduced earlier.

\vspace{-1.5mm}
\begin{minted}{haskell}
equivalence = do
    t1 <- forall "t1"
    t2 <- forall "t2"
    p1 <- forall "p1"
    p2 <- forall "p2"
    constrain $ t1 .>= 0 &&& t1 .<= toMilliSeconds (30 :: Year)
    constrain $ t2 .>= 0 &&& t2 .<= toMilliSeconds (30 :: Year)
    constrain $ p1 .>= 0 &&& p1 .<= toMilliWatts   ( 1 :: Watt)
    constrain $ p2 .>= 0 &&& p2 .<= toMilliWatts   ( 1 :: Watt)
    let dataMemory   = [t1, t2, p1, p2, 0, 100]
        llFinalState = simulate 100 (boot energyEstimateLowLevel  dataMemory)
        hlFinalState = simulate 100 (boot energyEstimateHighLevel dataMemory)
        llResult     = readArray (registers llFinalState) r0
        hlResult     = readArray (registers hlFinalState) r0
    return $ llResult .== hlResult
\end{minted}
\vspace{-1.5mm}

\noindent
We run Z3 (now on 52 SMT clauses) and get the affirmative result in 11.5s:

\vspace{-1mm}
\begin{minted}[frame=single]{text}
> proveWith z3 equivalence
Q.E.D.
\end{minted}
\vspace{-1mm}

\vspace{-3mm}
\subsubsection{Program timing analysis}

The REDFIN semantic core implements the system clock tracking with the~\hs{delay}
function:

\vspace{-1mm}
\begin{minted}{haskell}
delay :: Clock -> Redfin ()
delay cycles = Redfin $ \s -> ((), s { clock = clock s + cycles })
\end{minted}

Every execution step~-- a call of the~\hs{executeInstruction} function~-- advances
the clock by the number of cycles required to do the associated computation as well
as memory and register accesses. The tracking is precisely matched to the hardware
implementation and enables the engineers to
perform~\emph{best/worst case execution timing analysis} exploiting the
optimisation facilities provided by SBV and Z3. As an example, let us determine
the minimum and maximum number of clock cycles required for
executing~\hs{energyEstimateLowLevel}. For the sake of making this example more
interesting, we modified the semantics of the instruction~\hs{abs} and
added~1~extra clock cycle in case of a negative argument by conditionally performing
\hs{delay 1} in the state transformer \hs{abs}.

\vspace{-1mm}
\begin{minted}{haskell}
timingAnalysis = optimize Independent $ do
    t1 <- exists "t1"
    t2 <- exists "t2"
    p1 <- exists "p1"
    p2 <- exists "p2"
    constrain $ t1 .>= 0 &&& t1 .<= toMilliSeconds (30 :: Year)
    constrain $ t2 .>= 0 &&& t2 .<= toMilliSeconds (30 :: Year)
    constrain $ p1 .>= 0 &&& p1 .<= toMilliWatts   ( 1 :: Watt)
    constrain $ p2 .>= 0 &&& p2 .<= toMilliWatts   ( 1 :: Watt)
    let dataMemory = [t1, t2, p1, p2, 0, 100]
        finalState = simulate 100 (boot energyEstimateLowLevel dataMemory)
    -- Specify independent optimisation goals:
    minimize "Best case"  (clock finalState)
    maximize "Worst case" (clock finalState)
\end{minted}
\vspace{-1mm}

\noindent
The total delay of the program depends only on the sign of $t_1 - t_2$, thus
the best and worst cases differ only by one clock cycle. The worst case is
achieved when the difference is negative ($t_1 - t_2 = -2$), as shown below.
Z3 finishes in 0.5s.

\vspace{2mm}
\begin{minipage}{0.45\textwidth}
\begin{minted}[frame=single]{text}
Objective "Best case":
Optimal model:
  t1        = 549755813888
  t2        =  17179869184
  p1        =            0
  p2        =            0
  Best case =           12
\end{minted}
\end{minipage}
\begin{minipage}{0.45\textwidth}
\begin{minted}[frame=single]{text}
Objective "Worst case":
Optimal model:
  t1         = 65535
  t2         = 65537
  p1         =     0
  p2         =     0
  Worst case =    13
\end{minted}
\end{minipage}
\vspace{-2mm}

\vspace{1mm}
\section{Discussion\label{sec-discussion}}
\vspace{-3mm}
The presented approach has been implemented and is planned to be released at the
conference (this paper is currently under review).
In this section we discuss our main design choices and achieved
results, comparing them with the project's initial goals: (i) providing a unified
specification, testing, and formal verification framework that is (ii) understandable
and convenient to use by the REDFIN engineering team, and (iii)~allows the team
to co-develop REDFIN software and hardware, by extending and modifying the
default instruction semantics.

\subsection{From hardware to untyped assembly to typed software}
\vspace{-1mm}
The proposed approach covers two levels of organisation of computer systems: the
hardware microarchitecture (the state monad \hs{Redfin},~\S\ref{sec-transformer}),
and the instruction set architecture (the assembly monad
\hs{Script},~\S\ref{sec-verification}). These two levels are very different: the
former allows hardware engineers to precisely capture the program semantics (and
has proven useful in exposing underspecified behaviours), whereas the latter does
not have a direct access to the microarchitectural level and is used to
symbolically \emph{execute} the semantics, allowing software engineers to
\emph{observe} the results and \emph{reason} about program correctness.

By using Haskell as a metalanguage, we provided a purely syntactic implementation
of a higher-level abstraction on top of the REDFIN assembly~---~a statically-typed
language for arithmetic expressions. This demonstrates that the user
of the verification framework has enough power to implement their own domain-specific
extensions of the REDFIN assembly.

Although our current implementation is written in Haskell, the presented ideas
can be implemented in many other languages. We chose Haskell for its built-in
support for polymorphic expressions, powerful \hs{do}-notation, and availability
of a mature symbolic manipulation library (SBV). We also have a prototype
implementation in Idris, a much younger language that features dependent
types and hence allows us to verify more sophisticated properties at the type
level~\cite{JFP:9060502}, however at the time of writing there is no equivalent
of the SBV library in Idris, which is a significant practical disadvantage.
Dependent Haskell~\cite{weirich2017dependent}, once implemented, will provide
a convenient alternative.

\vspace{-2.5mm}
\subsection{Uniform development, testing and verification environment}
\vspace{-0.5mm}
The \hs{Script} monad was engineered to provide familiar assembly mnemonics and
directives (e.g. data and instruction labels), which allows engineers to start
using the framework for developing REDFIN programs even without prior experience
of Haskell development, hopefully increasing the uptake of the framework.

Thanks to symbolic simulation, we can uniformly handle both concrete and
symbolic values, thus testing becomes just a special case of formal verification,
allowing the engineers to reuse a common code base and infrastructure.
Testing yields trivial SMT problems that can be solved in sub-second time for
all programs of realistic sizes (typical REDFIN programs have hundreds of
instructions). Formal verification is more expensive: in our experiments, we
could handle programs comprising hundreds of instructions in 10-15 minutes, but
one can easily construct small programs that will grind any SMT solver to a halt:
for example, analysis of a single multiplication instruction \hs{mul} can take
half an hour if it is required to factor 64"/bit numbers (try factoring
\hs{4611686585363088391} with an SMT solver). In such cases, conservatively proving
some of the correctness properties at the type level can significantly increase
the productivity.

Although proving properties about the hardware implementation is left for future
work, the developed infrastructure provides a way to generate testsuites for the
processing core from the formal semantics of REDFIN instructions. Furthermore,
one can use the semantics to generate parts of the hardware
implementation~\cite{reid2016cav} or synthesise efficient instruction
subsets~\cite{mokhov2014synthesis}.

\section{Related Work\label{sec-related}}
\vspace{-1mm}
There is a vast body of literature available on the topic of formal verification,
including verification of hardware processing cores and low-level software programs.
Our work builds in a substantial way on a few known ideas that we will review in
this section. We thank the formal verification and programming languages
communities and hope that the formal semantics of the REDFIN processing core will
provide a new interesting benchmark for future studies.

We model the REDFIN microarchitecture using a~\emph{monadic state transformer
metalanguage}. There are several examples of prior work exploiting this idea.
Fox and Myreen~\cite{fox2010trustworthy} formalise the Arm v7 instruction
set architecture in HOL4 and give a careful account to bit-accurate proofs of
the instruction decoder correctness. Later, Kennedy et al.~\cite{kennedy2013coq}
formalised a subset of the x86 architecture in Coq, using monads for instruction
execution semantics and monadic \hs{do}-notation for assembly language embedding.
Both these models are formalised in proof assistants, thus are powered by full
dependent types, which allow the usage of mechanised program correctness proofs.
Degenbaev~\cite{degenbaev2012formal} formally specifies the \emph{complete} x86
instruction set -- a truly monumental effort! -- using a custom domain-specific
language that can be translated to a formal proof system. Arm's Architecture
Specification Language has been developed for the same purpose to formalise the
Arm v8 instruction set~\cite{reid2016cav}. Our specification approach is similar
to these two works, but we operate on a much smaller scale of the REDFIN core.

Our monadic metalanguage is embedded in Haskell and does not have a rigorous
formalisation, i.e. we cannot prove correctness of the REDFIN semantics
itself (this is a common concern, e.g. see~\cite{reid2017oopsla}). Moreover, our
verification workflow mainly relies on \emph{automated} theorem proving, rather
than on \emph{interactive} one. This is motivated by the cost of precise proof
assistant formalisations in terms of human resources: automated techniques are
more CPU-intensive, but cause less ``human-scaling issues''~(Reid at
al.~\cite{reid2016cav}). Our goal was to create a framework that could be seamlessly
integrated into an existing spacecraft engineering workflow, therefore it needed
to have as much proof automation as possible. The automation is achieved by means
of \emph{symbolic program execution}. Currie at al.~\cite{Currie2006} applied
symbolic execution with uninterpreted functions to prove equivalence of low-level
assembly programs. The framework we present allows not only proving the
equivalence of low-level programs, but also their compliance with higher-level
specifications written in a subset of Haskell.

A lot of research work has been done on the design of \emph{typed assembly
languages}, e.g. see~\cite{Haas:2017:BWU:3140587.3062363}\cite{Morrisett:1999:SFT:319301.319345}.
The low-level REDFIN assembly is untyped, but the syntactic language of
arithmetic expressions that we implemented on top of it does have a simple type
system. In principle, the REDFIN assembly itself may benefit from a richer type
system, especially one enforcing correct operation with relevant mission-specific
units of measurement~\cite{Kennedy:1997:RPU:263699.263761}.

Finally, we would like to acknowledge several related projects, blogposts and talks
that provided an initial inspiration for this work: the `Monads to Machine
Code' compiler by Diehl~\cite{diehl-monads-to-machines}, RISC-V semantics in Haskell
by MIT~\cite{riscv-semantics}, Wall's assembly monad~\cite{asm-monad}, and
SMT-based program analysis by Jelvis~\cite{haskell-z3}.

\section{Conclusions and opportunities for future research\label{sec-future}}
This paper presents a formal verification approach developed to tackle a real
engineering problem by combining known techniques from the formal verification
and programming languages research communities. We demonstrated that these
techniques can be applied to create a formal model of the spacecraft processing
core REDFIN that is used to execute small control programs, and showed how to
formally verify such programs. By releasing the state transformer semantics to
public, we hope to provide other researchers with a realistic benchmark for
their formal verification tools. We also invite them to contribute to the
REDFIN toolchain itself. Below we list opportunities for future research.

\subsection{Dependently-typed high-level DSL for REDFIN programs}

On top of low-level REDFIN assembly we implemented a high-level arithmetic
expression language with a simple type system to distinguish between variables
using different number representation and measurement units. This very simple
type system already helps to eliminate an important class of bugs
with the help of Haskell's type checker. However, a dependently-typed host language,
such as Agda, Coq, Dependent Haskell, or Idris, could provide us much more power.
One example is compile-time elimination of out-of-memory access that can occur
when branching to a non-existent program location or overflowing the stack.

\subsection{System-level verification via strongly typed protocols}

The REDFIN instruction set architecture has a number of bus-communication instructions
to access the system bus. An interesting research problem is implementing a
model of a complete space system in an advanced typed programming language by modelling
each component separately and later integrating them using shared types specifying
the communication protocol. We can then derive bus-communication code
from the protocol specification, while sharing the same types on both sides will
allow ensuring the correctness of communication. Similar work has been
done in the context of web systems~\cite{Mestanogullari:2015:TWA:2808098.2808099}.

\subsection{Hardware synthesis}

The state transformer metalanguage presented in the paper makes the system
state explicit and employs Haskell's advanced abstractions to make the
state manipulation safe and structured. To convey the model's verification
power down to the bare metal, we can implement a verified translation from the
state-transformer metalanguage to a hardware description language to petrify
the semantics and turn it into silicon.


\section*{Acknowledgements}
We would like to thank Neil Mitchell, Charles Morisset, Artem Pelenitsyn and
Danil Sokolov for their helpful feedback on an earlier version of this paper.

\bibliography{biblio}

\begin{thebibliography}{10}
\providecommand{\url}[1]{\texttt{#1}}
\providecommand{\urlprefix}{URL }

\bibitem{SBV}
{SBV: SMT Based Verification in Haskell},
  \url{http://leventerkok.github.io/sbv/}

\bibitem{NASA:1999:Mars}
{Mars Climate Orbiter Mishap Investigation Board Phase I Report}. Tech. rep.,
  NASA (Nov 1999)

\bibitem{bug-rocket}
Ben-Ari, M.: {The Bug That Destroyed a Rocket}. SIGCSE Bull.  33(2),  58--59
  (June 2001)

\bibitem{JFP:9060502}
Brady, E.: {Idris, a general-purpose dependently typed programming language:
  Design and implementation}. Journal of Functional Programming  23,  552--593
  (9 2013)

\bibitem{Currie2006}
Currie, D., Feng, X., Fujita, M., Hu, A.J., Kwan, M., Rajan, S.: {Embedded
  Software Verification Using Symbolic Execution and Uninterpreted Functions}.
  International Journal of Parallel Programming  34(1),  61--91 (2006)

\bibitem{de2008z3}
De~Moura, L., Bj{\o}rner, N.: {Z3: An efficient SMT solver}. Tools and
  Algorithms for the Construction and Analysis of Systems pp. 337--340 (2008)

\bibitem{degenbaev2012formal}
Degenbaev, U.: Formal specification of the x86 instruction set architecture.
  Ph.D. thesis, Saarland University (2012)

\bibitem{diehl-monads-to-machines}
Diehl, S.: {Monads to Machine Code}.
  \url{http://www.stephendiehl.com/posts/monads_machine_code.html} (2017)

\bibitem{fox2010trustworthy}
Fox, A., Myreen, M.O.: {A trustworthy monadic formalization of the ARMv7
  instruction set architecture}. In: International Conference on Interactive
  Theorem Proving. pp. 243--258. Springer (2010)

\bibitem{Haas:2017:BWU:3140587.3062363}
Haas, A., Rossberg, A., Schuff, D.L., Titzer, B.L., Holman, M., Gohman, D.,
  Wagner, L., Zakai, A., Bastien, J.: {Bringing the Web Up to Speed with
  WebAssembly}. SIGPLAN Not.  52(6),  185--200 (2017)

\bibitem{haskell-z3}
Jelvis, T.: {Analyzing Programs with Z3 (video recording of Compose Conference
  talk)}. http://jelv.is/talks/compose-2016 (2016)

\bibitem{kennedy2013coq}
Kennedy, A., Benton, N., Jensen, J.B., Dagand, P.E.: Coq: the world's best
  macro assembler? In: Proceedings of the 15th Symposium on Principles and
  Practice of Declarative Programming. pp. 13--24. ACM (2013)

\bibitem{Kennedy:1997:RPU:263699.263761}
Kennedy, A.J.: {Relational Parametricity and Units of Measure}. In: Proceedings
  of the 24th ACM SIGPLAN-SIGACT Symposium on Principles of Programming
  Languages. pp. 442--455. POPL'97, ACM (1997)

\bibitem{Leveson2004}
Leveson, N.G.: {Role of Software in Spacecraft Accidents}. Journal of
  Spacecraft and Rockets  41(4),  564--575 (2004)

\bibitem{Mestanogullari:2015:TWA:2808098.2808099}
Mestanogullari, A., Hahn, S., Arni, J.K., L\"{o}h, A.: {Type-level Web APIs
  with Servant: An Exercise in Domain-specific Generic Programming}. In:
  Proceedings of the 11th ACM SIGPLAN Workshop on Generic Programming. pp.
  1--12. WGP 2015, ACM (2015)

\bibitem{riscv-semantics}
{MIT CSAIL}: {A formal specification of the RISC-V ISA written in Haskell}.
  \url{https://github.com/mit-plv/riscv-semantics} (2017)

\bibitem{mokhov2014synthesis}
Mokhov, A., Iliasov, A., Sokolov, D., Rykunov, M., Yakovlev, A., Romanovsky,
  A.: {Synthesis of processor instruction sets from high-level ISA
  specifications}. IEEE Transactions on Computers  63(6),  1552--1566 (2014)

\bibitem{Morrisett:1999:SFT:319301.319345}
Morrisett, G., Walker, D., Crary, K., Glew, N.: {From System F to Typed
  Assembly Language}. ACM Trans. Program. Lang. Syst.  21(3),  527--568 (1999)

\bibitem{reid2017oopsla}
Reid, A.: {Who Guards the Guards? Formal Validation of the Arm V8-m
  Architecture Specification}. Proc. ACM Program. Lang.  1(OOPSLA),
  88:1--88:24 (2017)

\bibitem{reid2016cav}
Reid, A., Chen, R., Deligiannis, A., Gilday, D., Hoyes, D., Keen, W.,
  Pathirane, A., Shepherd, O., Vrabel, P., Zaidi, A.: {End-to-end verification
  of processors with ISA-Formal}. In: International Conference on Computer
  Aided Verification. pp. 42--58. Springer (2016)

\bibitem{vazou2014refinement}
Vazou, N., Seidel, E.L., Jhala, R., Vytiniotis, D., Peyton-Jones, S.:
  {Refinement types for Haskell}. In: ACM SIGPLAN Notices. vol.~49, pp.
  269--282. ACM (2014)

\bibitem{wadler1990comprehending}
Wadler, P.: Comprehending monads. In: Proceedings of the 1990 ACM conference on
  LISP and functional programming. pp. 61--78. ACM (1990)

\bibitem{asm-monad}
Wall, L.: {An ASM Monad}.
  \url{http://wall.org/~lewis/2013/10/15/asm-monad.html} (2017)

\bibitem{weirich2017dependent}
Weirich, S., Voizard, A., de~Amorim, P.H.A., Eisenberg, R.A.: {A specification
  for dependent types in Haskell}. Proceedings of the ACM on Programming
  Languages  1(ICFP), ~31 (2017)

\end{thebibliography}
\end{document}